\documentclass[12pt]{article}

\usepackage{epsfig}
\usepackage{citesort}

\def  \bcen   {\begin{center}}
\def  \ecen   {\end{center}}
\def  \beq    {\begin{equation}}
\def  \eeq    {\end{equation}}
\def  \beqa   {\begin{eqnarray}}
\def  \eeqa   {\end{eqnarray}}

\def  \mks    {m_{K^*}}
\def  \eks    {E_{K^*}} 

\def  \unpart {{\cal U}} 
\def  \du     {{d_{\cal U}}}
\def  \ou     {O_{\cal U}}
\def  \lamu   {\Lambda_{\cal U}} 
\def  \MU     {M_{\unpart}}

\def  \cs     {{\cal C}_S}
\def  \cp     {{\cal C}_P} 
\def  \cv     {{\cal C}_V}
\def  \ca     {{\cal C}_A}
\def  \calm   {{\cal M}} 

\def  \btsnn  {b \to s \nu \bar{\nu}}
 
\def  \bkksnn {B \to K(K^*) \nu \bar{\nu} }
\def  \bkksu  {B \to K(K^*) \ {\cal U} }
\def  \miss   {{\rm missing \ energy}} 
\def  \misse  {{\displaystyle{\not} E}}

\begin{document}

\renewcommand{\thefootnote}{\fnsymbol{footnote}}

\begin{flushright}
KEK-TH-1153, \\
LYCEN 2007-05.
\end{flushright}
\vskip 2cm
\bcen
{\large \bf \boldmath $B \to K(K^*) + {\rm missing \ energy}$ in Unparticle physics}  
\vskip 1.8cm
{\sc \large T. M. Aliev$^1$\footnote{taliev@metu.edu.tr}, 
A. S. Cornell$^2$\footnote{cornell@ipnl.in2p3.fr} and 
Naveen Gaur$^3$\footnote{naveen@post.kek.jp} }
\vskip .7cm
{\sl $^1$ Physics Department, Middle East Technical University, 06531 Ankara, Turkey, \\
$^2$ Universit\'e de Lyon 1, Institut de Physique Nucl\'eaire, Villeurbanne, France, \\ 
$^3$ Theory Division, KEK, 1-1 Oho, Tsukuba, Ibaraki 305-0801, Japan}
\ecen
\vskip 1.8cm


\thispagestyle{empty}

\begin{abstract}
\par In the present work we study the effects of an unparticle
$\unpart$ as the possible source of missing energy in the decay $B \to
K (K^*) + {\rm missing \ energy}$. We find that the dependence of the
differential branching ratio on the $K$($K^*$)-meson's energy in the
presence of the vector unparticle operators is very distinctive from
that of the SM. Moreover, in using the existing upper bound on $B \to
K (K^*) + {\rm missing \ energy}$ decays, we have been able to put
more stringent constraints on the parameters of unparticle stuff.  
\end{abstract}

\vfill \eject 


\section{Introduction}

\par Flavour Changing Neutral Current (FCNC) processes are not only
powerful tests of the Standard Model (SM) but also provide very
stringent tests for any physics beyond it. The smallness of FCNC
processes in the SM is attributed to the fact that these processes are
generated at loop level and are further suppressed by the CKM
factors. Due to their smallness within the SM these processes can also
be very sensitive to any new physics beyond the SM. Amongst the many
FCNC decays involving $B$ and $K$-mesons the decays of the form $b \to
s + {\rm missing \ energy}$ have been the focus of much investigation
at the $B$ factories Belle and Babar.   

\par Of particular interest, in the SM, is the decay $b \to s \nu
\bar{\nu}$, as it has the theoretical advantage of uncertainties much
smaller than those of other decays, due to the absence of a photonic
penguin contribution and hadronic long distance effects. However, in
spite these theoretical advantages, it might be very difficult to
measure the inclusive mode $B \to X_s \nu \bar{\nu}$, as it requires a
construction of all the $X_s$'s. Therefore the rare $B \to K (K^*) 
\nu \bar{\nu}$ decays play a special role, both from experimental and
theoretical points of view. Also the branching fractions of the
$B$-meson decays are quite large, with theoretical estimates of $Br(B
\to K^* \nu \bar{\nu}) \sim 10^{-5}$ and $Br(B \to K \nu \bar{\nu})
\sim 10^{-6}$ \cite{Buchalla:2000sk}. These processes, based on
$\btsnn$, are very sensitive to non-standard $Z$ models and have been
extensively studied in the literature
\cite{Kim:1999wa,Bird:2004ts,Aliev:2001in}. 

\par As such, any new physics model which can provide a relatively
light new source of {\it missing energy} can potentially enhance the
observed rates of $B \to K (K^*) + {\rm missing \ energy}$ ($B \to K
(K^*) + \misse$), where many models have been proposed which provide
such low mass candidates (which can contribute to $b \to s
\miss$). Note that in reference \cite{Bird:2004ts} the phenomenology
of such low mass scalars was explored. Such studies have also been
done in the context of large extra dimension models
\cite{Mahajan:2003gx} and leptophobic $Z'$ models
\cite{Buchalla:2000sk,Kim:1999wa}. One such model, which has excited
much interest recently, is that of Unparticles, as proposed by
H. Georgi \cite{Georgi:2007ek}. In this model we assume that at a very
high energy our theory contains both the fields of the SM and the
fields of a theory with a nontrivial IR fixed point, which he called
the Banks-Zaks (BZ) fields \cite{Banks:1981nn}. In his model these two
sets interacted through the exchange of particles with a large mass
scale $\MU$, where below this scale there were nonrenormablizable
couplings between the SM fields and the BZ fields suppressed by powers
of $\MU$. The renormalizable couplings of the BZ fields then produced
dimensional transmutation, and the scale-invariant unparticle fields
emerged below an energy scale $\lamu$. 

\par In the effective theory below $\lamu$ the BZ operators matched
onto the unparticle operators, and the nonrenormaliable interactions
matched onto a new set of interactions between the SM and unparticle
fields. The end result was a collection of unparticle stuff with scale
dimension $\du$, which looked like a non-integral number $\du$ of
invisible massless particles, whose production might be detectable in
missing energy and momentum distributions \cite{Georgi:2007ek}.   

\par Recently there has been a lot of interest in unparticle physics
\cite{Georgi:2007ek,Georgi:2007si,Luo:2007bq,Cheung:2007ue,Liao:2007bx,Li:2007by,Choudhury:2007js,Stephanov:2007ry,Fox:2007sy,Davoudiasl:2007jr},
where the signatures of {\sl unparticles} have been discussed at
colliders \cite{Georgi:2007si,Cheung:2007ue,Fox:2007sy}, in Lepton
Flavor Violating (LFV) processes \cite{Choudhury:2007js}, cosmology
and astrophysics \cite{Davoudiasl:2007jr}, and low energy processes
\cite{Liao:2007bx,Li:2007by,Luo:2007bq,Li:2007by}. 

\par In the present work we study the $B \to K (K^*) + \misse$ decay
in unparticle theory, where this work is organized as follows: In
section 2 we calculate the various contributions from both the SM and
unparticle theory to the above-mentioned decays. Section 3 contains our
numerical analysis and conclusions.   


\section{Differential Decay Widths}

\par In the SM the decay mode $B \to K(K^*) + \misse$ is described by
the decay $B \to K (K^*) \nu \bar{\nu}$. As was noted earlier,
unparticles can also contribute to these decays. Hence a comparison of
the signatures of the two decay modes $B \to K (K^*) \nu \bar{\nu}$
and $B \to K (K^*) \unpart$ is required. 

\par In the SM the decay $B \to K (K^*) \nu \bar{\nu}$ is described by
the quark level process $b \to s \nu \bar{\nu}$ through the effective
Hamiltonian:  
\beq
{\cal H} = \frac{G_F}{\sqrt{2}} \frac{\alpha}{2 \pi} V_{tb}
V^*_{ts}C_{10}  
~ \bar{s} \gamma_\mu \left(1 - \gamma_5 \right) b ~ 
\bar{\nu} \gamma_\mu \left(1 - \gamma_5\right) \nu \,\,\, ,
\label{eq:sec2:1}
\eeq
where
\beq
C_{10} = \frac{X(x_t)}{sin^2\theta_w} \,\,\, ,
\label{eq:sec2:2}
\eeq
and the $X(x_t)$ is the usual Inami-Lim function, given as:
\beq
X(x_t) = \frac{x_t}{8} \left\{ \frac{x_t + 1}{x_t - 1} + \frac{3 x_t -
    6}{(x_t - 1)^2} ln(x_t) \right\} \,\,\, , 
\label{eq:sec2:3} 
\eeq
with $x_t = m_t/m_W^2$. 

\par Similarly, the unparticle transition at quark level can be
described by $b \to s \unpart$, where we shall consider the following
operators: 
\beqa
{\rm Scalar \ unparticle \ operators} \ \ &\Longrightarrow & \ \ \cs
\frac{1}{\lamu^{\du}} ~ \bar{s} \gamma_\mu b ~ \partial^\mu\ou + \cp
\frac{1}{\lamu^{\du}} ~ \bar{s} \gamma_\mu \gamma_5 b
~ \partial^\mu\ou  \,\,\, , \nonumber \\   
{\rm Vector \ unparticle \ operators} \ \ &\Longrightarrow & \ \ \cv
\frac{1}{\lamu^{\du - 1}} ~ \bar{s} \gamma_\mu b ~ \ou^\mu  \ + \ \ca
\frac{1}{\lamu^{\du - 1}} ~ \bar{s} \gamma_\mu \gamma_5 b ~ \ou^\mu
\,\,\, . 
\label{eq:sec2:4}  
\eeqa

\par Before proceeding with our analysis note that we shall write the
propagator for the scalar unparticle field as
\cite{Georgi:2007si,Cheung:2007ue}:   
\beq
\int d^4 x e^{iP.x} \langle 0|T \ou (x) \ou (0) | 0 \rangle 
= i \frac{A_{\du}}{2 ~ {\rm sin}(\du \pi)} (- P^2)^{\du - 2} \,\,\, , 
\label{eq:sec2:5}
\eeq
where
$$
A_{\du} = \frac{16 \pi^{5/2}}{(2 \pi)^{2 \du}} 
\frac{\Gamma(\du + 1/2)}{\Gamma(\du - 1) \Gamma(2\du)} \,\,\, .
$$


\subsection{The Standard Model} 

\par Using the SM effective Hamiltonian for the quark level process $b
\to s \nu \bar{\nu}$, as given in equation (\ref{eq:sec2:1}), we can
calculate the differential decay width of $\bkksnn$ (using the form
factor definitions for the $B \to K$ transition as given in appendix
\ref{appendix:a1}).  

\par After taking into account the three species of SM neutrinos, we
evaluate the differential decay width as a function of $K$-meson
energy ($E_K$) as: 
\beq
\frac{d \Gamma^{SM}}{d E_K} = \frac{G_F^2 \alpha^2}{2^7 \pi^5 m_B^2} ~
|V_{ts} V_{tb}^*|^2 ~ |C_{10}|^2 f_+^2(q^2) ~
\lambda^{3/2}(m_B^2,m_K^2,q^2)   \,\,\, ,  
\label{eq:sec2:6}
\eeq
where $\lambda(m_B^2,m_K^2,q^2) = m_B^4 + m_K^4 + q^4 - 2 m_B^2 q^2 -
2 m_K^2 q^2 - 2 m_K^2 m_B^2$, and $q^2 = m_B^2 + m_K^2 - 2 m_B E_K$.  

\par Similarly, for the $B \to K^*$ case, using the definition of form
factors for $B \to K^*$ transitions as given in appendix
\ref{appendix:a2}, the differential decay rate in the SM can be
calculated as: 
\beqa
\frac{d\Gamma^{SM}}{d E_{K^*}} = \frac{G_F^2 \alpha^2}{2^9 \pi^5 m_B^2} 
|V_{ts} V_{tb}^*|^2 \lambda^{1/2} |C_{10}|^2 
\bigg( 8 \lambda q^2 \frac{V^2}{(m_B + m_{K^*})^2} 
+ \frac{1}{m_{K^*}^2} 
\bigg[ \lambda^2 \frac{A_2^2}{(m_B + m_{K^*})^2}  \nonumber \\
+ (m_B + m_{K^*})^2 (\lambda + 12 m_{K^*}^2 q^2 ) A_1^2 
- 2 \lambda (m_B^2 - m_{K^*}^2 - q^2) Re(A_1^* A_2) 
\bigg]
\bigg) \,\,\, ,
\label{eq:sec2:7}
\eeqa
where $\lambda = m_B^4 + m_{K^*}^4 + q^4 - 2 m_B^2 q^2 - 2 m_{K^*}^2
q^2 - 2 m_{K^*}^2 m_B^2$, and $q^2 = m_B^2 + m_{K^*}^2 - 2 m_B \eks$.   


\subsection{The Scalar Unparticle Operator}

\par As listed earlier, the following scalar operators can contribute
to the $\bkksu$ decay: 
\beq
\cs \frac{1}{\lamu^{\du}} ~ \bar{s} \gamma_\mu b ~ \partial^\mu\ou +
\cp \frac{1}{\lamu^{\du}} ~ \bar{s} \gamma_\mu \gamma_5 b
~ \partial^\mu\ou = \frac{1}{\lamu^{\du}} ~ \bar{s} \gamma_\mu
\left(\cs + \cp \gamma_5\right) b ~ \partial^\mu\ou \,\,\,
, \label{eq:sec2:8} 
\eeq
where we have defined our form factors in appendix A. As such, the
matrix element for the process $B(p) \to K(p') + \unpart(q)$ can be
written as:  
\beq
\calm^S  =  \frac{1}{\lamu^{\du}} ~ \cs \bigg[ f_+ (m_B^2 - m_K^2) + f_- q^2
 \bigg] ~ \ou \,\,\, . 
\label{eq:sec2:9} 
\eeq
The decay rate for $B(p) \to K(p') \unpart(q)$ can now be evaluated to be:
\beqa
\frac{d\Gamma^{S \unpart}}{d E_K} &=& \frac{1}{8 \pi^2 m_B} 
\frac{A_\du}{\lamu^{2\du}} ~
|\cs|^2 \sqrt{E_K^2 - m_K^2} \left(m_B^2 + m_K^2 - 2 m_B
  E_K\right)^{\du - 2}        \nonumber \\
&& \hskip 1in \times \bigg[ f_+ (m_B^2 - m_K^2) + f_- (m_B^2 + 2 m_K^2
- 2 m_B E_K) \bigg]^2 \,\,\, . 
\label{eq:sec2:10}
\eeqa

\par For the $B \to K^*$ transition our calculation proceeds along the
same lines, where the matrix element for $B(p) \to K^*(p') \unpart(q)$
can be written as: 
\beq
\calm^S = \frac{i \cp}{\lamu^{\du}} (\epsilon . q) \left\{ (m_B + m_{K^*}) A_1 - (m_B -
  m_{K^*}) A_2 - 2 m_{K^*} \left( A_3 - A_0 \right) \right\} ~ \ou \,\,\, ,  
\eeq
and the differential decay rate as:
\beq
\frac{d\Gamma^{S \unpart}}{d E_{K^*}} = \frac{m_B}{2 \pi^2} ~
\frac{A_\du}{\Lambda_\unpart^{2\du}} ~ |\cp|^2 A_0^2 ~ \left(E_{K^*}^2
  - m_{K^*}^2\right)^{3/2} \left(m_B^2 + m_{K^*}^2 - 2 m_B E_{K^*}
\right)^{\du - 2} \,\,\, .  
\label{eq:sec2:11}
\eeq

As can seen from the above expressions the scalar unparticle
contribution to the decay rate for $B \to K \unpart$ and $B \to K^*
\unpart$ will depend upon $\cs$ and $\cp$ respectively. This shall
allow us to place constraints upon $\cs$ and $\cp$ from these two
different decay modes. This issue shall be re-visited in the final
section of this paper. 


\subsection{The Vector Unparticle Operator}

\par Along similar lines as followed in the previous subsection, we
shall now make use of the vector unparticle operators:  
$$
\cv \frac{1}{\lamu^{\du - 1}} ~ \bar{s} \gamma_\mu b ~ \ou^\mu  \ + \
\ca \frac{1}{\lamu^{\du - 1}} ~ \bar{s} \gamma_\mu \gamma_5 b ~
\ou^\mu  = \frac{1}{\lamu^{\du - 1}} ~ \bar{s} \gamma_\mu  \left( \cv
  + \ca \gamma_5 \right) b ~ \ou^\mu \,\,\, ,  
$$
and the form factors of appendix A, to calculate the matrix element
for $B(p) \to K(p') \unpart(q)$:  
\beq
\calm^V = \frac{1}{\lamu^{\du - 1}} ~ \cv \bigg[ f_+ (p + p')_\mu +
f_- (p - p')_\mu \bigg] ~ \ou^\mu \,\,\, . 
\label{eq:sec2:12} 
\eeq
And as such we calculate the differential decay rate as:
\beqa
\frac{d\Gamma^{V\unpart}}{d E_K} &=& \frac{1}{8 \pi^2 m_B}
\frac{A_\du}{\lamu^{2\du - 2}} ~ |\cv|^2 |f_+|^2 \left(m_B^2 + m_K^2 -
  2 m_B  E_K\right)^{\du - 2} \sqrt{E_K^2 - m_K^2} \nonumber \\ 
& & \hskip 1in \times \bigg\{ - (m_B^2 + m_K^2 + 2 m_B E_K) +
\frac{(m_B^2 - m_K)^2}{(m_B^2 + m_K^2 - 2 m_B E_K)} \bigg\} \,\,\,
. \label{eq:sec2:14} 
\eeqa

\par For the $B \to K^*$ case the matrix element for $B(p) \to K^*(p') \unpart(q)$ is:
\beqa
\calm^V & = &  \left\{ \frac{\ca}{\lamu^{\du - 1}} \left( i \epsilon_\mu (m_B + m_{K^*}) A_1 - i (p + p')_\mu (\epsilon.q) \frac{A_2}{m_B + m_{K^*}} - i q_\mu (\epsilon . q) \frac{2 m_{K^*}}{q^2} 
[ A_3 - A_0 ] \right) \right. \nonumber \\ 
&& \hspace{1cm} \left. + \frac{\cv}{\lamu^{\du - 1}} \left(\frac{2 V}{m_B + m_{K^*}} \epsilon_{\mu\nu \rho \sigma} \epsilon^\nu p^\rho p'^\sigma \right) \right\} ~ \ou^\mu \,\,\, . 
\eeqa
And therefore the differential decay rate will be:
\beqa
\frac{d\Gamma^{V\unpart}}{dE_{K^*}} &=& \frac{1}{8 \pi^2 m_B} (q^2)^{\du - 2}
\sqrt{E_{K^*}^2 - m_{K^*}^2}   
    \frac{A_\du}{\left(\Lambda_\unpart^{\du-1}\right)^2} ~
\Bigg\{ 8 |C_V|^2 m_B^2 \left(E_{K^*}^2 - m_{K^*}^2\right) 
\frac{V^2}{(m_B + m_{K^*})^2}   \nonumber \\
&& + |C_A|^2 \frac{1}{\mks^2 (m_B + \mks)^2 q^2} 
\bigg[ 
 (m_B + m_{K^*})^4 (3 m_{K^*}^4 + 2 m_B^2 m_{K^*}^2 - 6 m_B
  m_{K^*}^2 E_{K^*} + m_B^2 E_{K^*}^2) A_1^2   \nonumber \\
&& + 4 m_B^4 (\eks^2 - \mks^2)^2 A_2^2
+ 4 (m_B + \mks)^2 (m_B \eks - \mks^2) (\mks^2 - \eks^2) m_B^2 A_1 A_2
\bigg]
\Bigg\} \,\,\, . 
\label{eq:sec2:15}
\eeqa

\par To obtain the total decay width for $B \to K \unpart$ we must
integrate over $E_K$ in the range $m_K < E_K < (m_B^2 + m_K^2)/2 m_B$,
whereas to obtain the total decay width for $B \to K^* \unpart$ we
must integrate over $\eks$ in the range $\mks < \eks < (m_B^2 +
\mks^2)/2 m_B$.  


\section{Numerical Results and Conclusions}

\noindent The total contribution to $B \to K(K^*) + \misse$ can be written as: 
\beq
\Gamma = \Gamma^{SM} + \Gamma^\unpart \,\,\, , \label{eq:res1}
\eeq
where the $\Gamma^{SM}$ and $\Gamma^\unpart$ are the SM and unparticle
contributions to the $B \to K(K^*) + \misse$ decay. And we should note
that in the SM the missing energy in the final state is attributed to
the presence of neutrinos. Hence the SM contribution to this process
is given by $B \to K (K^*) \nu \bar{\nu}$. In the present case this
signature can be mimicked by the process $\bkksu$, where we shall now
try to estimate the bounds on the unparticles from the experimental
constraints on missing energy signatures, as given by the
$B$-factories BELLE and BaBar \cite{explimits,Aubert:2004ws}:    
\beqa
Br(B \to K \nu \bar{\nu})   &<&  1.4 \times 10^{-5} \,\,\, , \nonumber \\
Br(B \to K^* \nu \bar{\nu}) &<&  1.4 \times 10^{-4} \,\,\, . \nonumber
\eeqa

\begin{figure}[tb]
\bcen
\epsfig{file=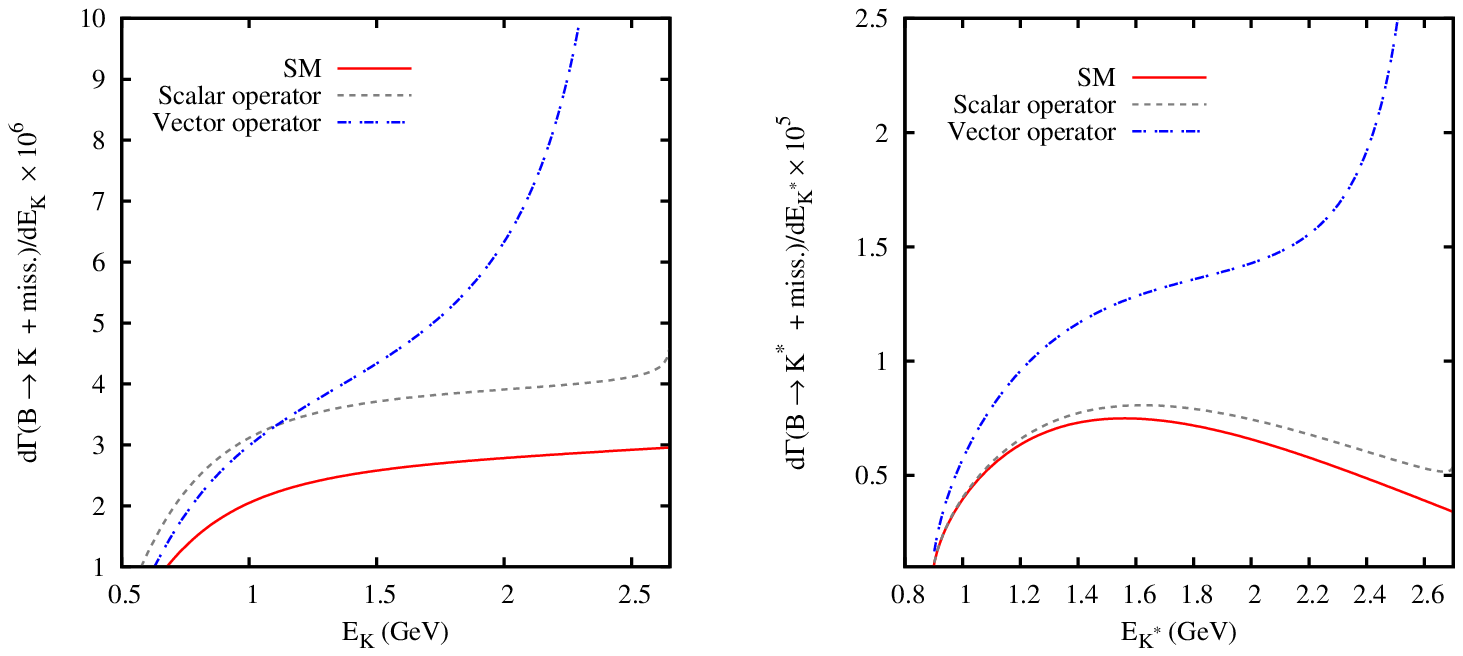,width=\textwidth}
\caption{\it The differential branching ratio for: (a) Left panel: $B
  \to K + \misse$ as a function of the hadronic energy ($E_K$). (b)
  Right panel: $B \to K^* + \misse$ as a function of the hadronic
  energy ($E_{K^*}$). The other parameters are $\du = 1.9$,
  $\Lambda_\unpart = 1000$GeV,  $\cp = \cs = 2 \times 10^{-3}$ and
  $\cv = \ca = 10^{-5}$.}    
\label{fig:1}
\ecen
\end{figure}

\par It is important to note that the SM process $\bkksnn$ provides a
unique energy distribution spectrum of final state hadrons ($K/K^*$ in
our case). Presently the experimental limits on the branching ratio of
these processes are about one order below the respective SM
expectation values. However, these processes are expected to be
measured at future SuperB factories. As such, we presently only have
an upper limit on the branching ratio of these processes, where to
estimate the constraints on the unparticle properties.  

\par Note that H. Georgi, in his first paper on unparticles, tried to
emphasize that unparticles behave as a non-integral number of
particles \cite{Georgi:2007ek}. He further went on to analyze the
distribution of the $u$-quark in the decay $t \to u \unpart$. It was
argued that the peculiar shape of the distributions of $E_u$ (the
energy of the $u$-quark) may allow us to discover unparticles
experimentally. As such, we have attempted to extend this same analogy
to the process presently under consideration. 

\par Finally, before presenting our numerical results, note that the
future SuperB factories will be measuring the process $B \to K(K^*) +
\misse$ by analyzing the spectra of the final state hadron. In doing
this measurement at $B$-factories a cut for high momentum on the
hadron is imposed, in order to suppress the background. Recall that
unparticles would give us an unique distribution for the high energy
hadron in the final state, such that in future $B$-factories one will
be able to distinguish the presence of a scale invariant sector (or
unparticles) by observing the spectrum of final state hadrons in $B
\to K(K^*) + \misse$.   

\par With this idea in mind we have tried to plot the differential
decay width of $B \to K(K^*) + \misse$ as a function of $E_K
(E_{K^*}$) in figure (\ref{fig:1}). As we can see from these figures
the unparticle operators (especially the vector operators) give us a
very distinctive distribution for the final state hadron's energy. The
distribution of the unparticle contribution is strikingly different
when we include a vector operator for a highly energetic final state
hadron. As such, unparticle stuff can give a distinctly different
signature from the SM in this regime, which it should be noted is
experimentally more favorable at future SuperB factories. 

\begin{figure}[tb]
\bcen
\epsfig{file=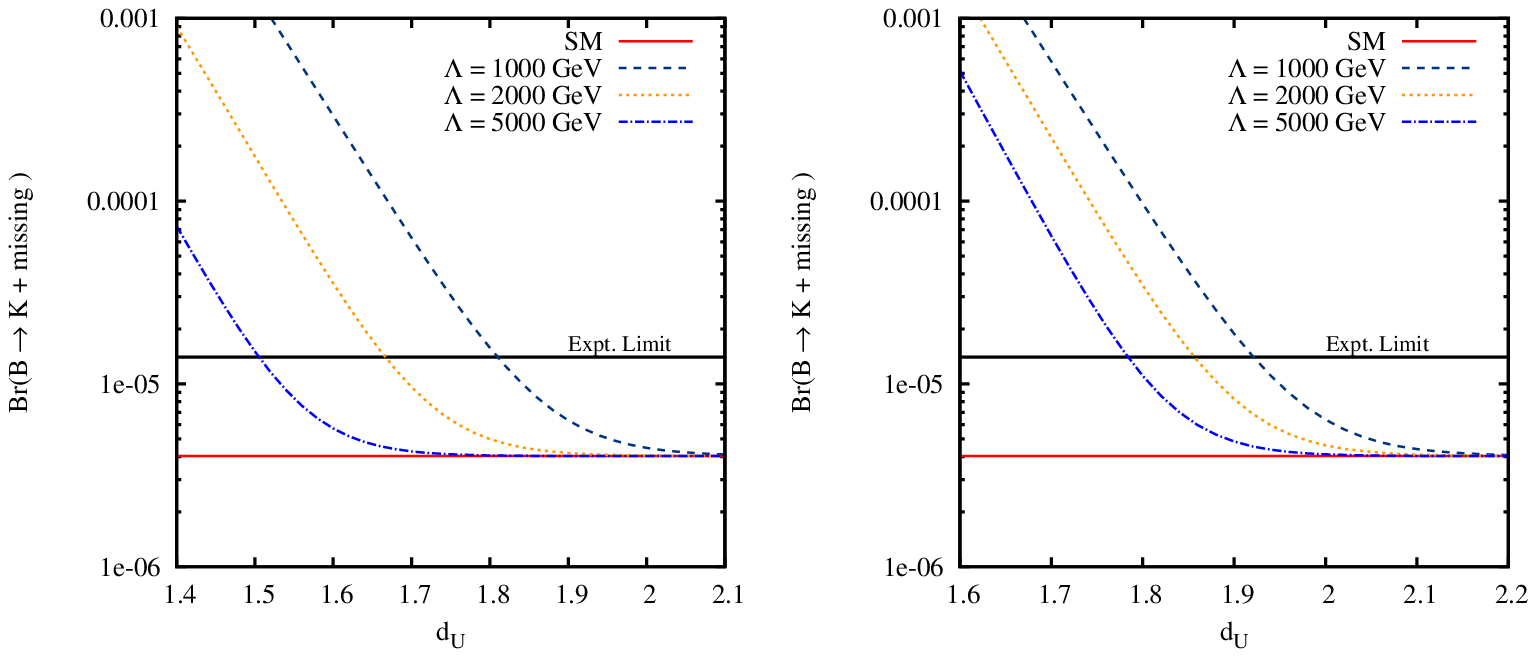,width=\textwidth} 
\caption{\it The branching ratio for $B \to K + \misse$ as a function
  of $\du$ for various values of $\Lambda_\unpart$. The left panel is
  for the contribution from the scalar operator, and the right panel
  is for the vector operator. The other parameters are $\cs = 2 \times
  10^{-3}$ and $\cv = 10^{-5}$.}   
\label{fig:2}
\ecen
\end{figure}

\par
\begin{figure}[htb]
\bcen
\epsfig{file=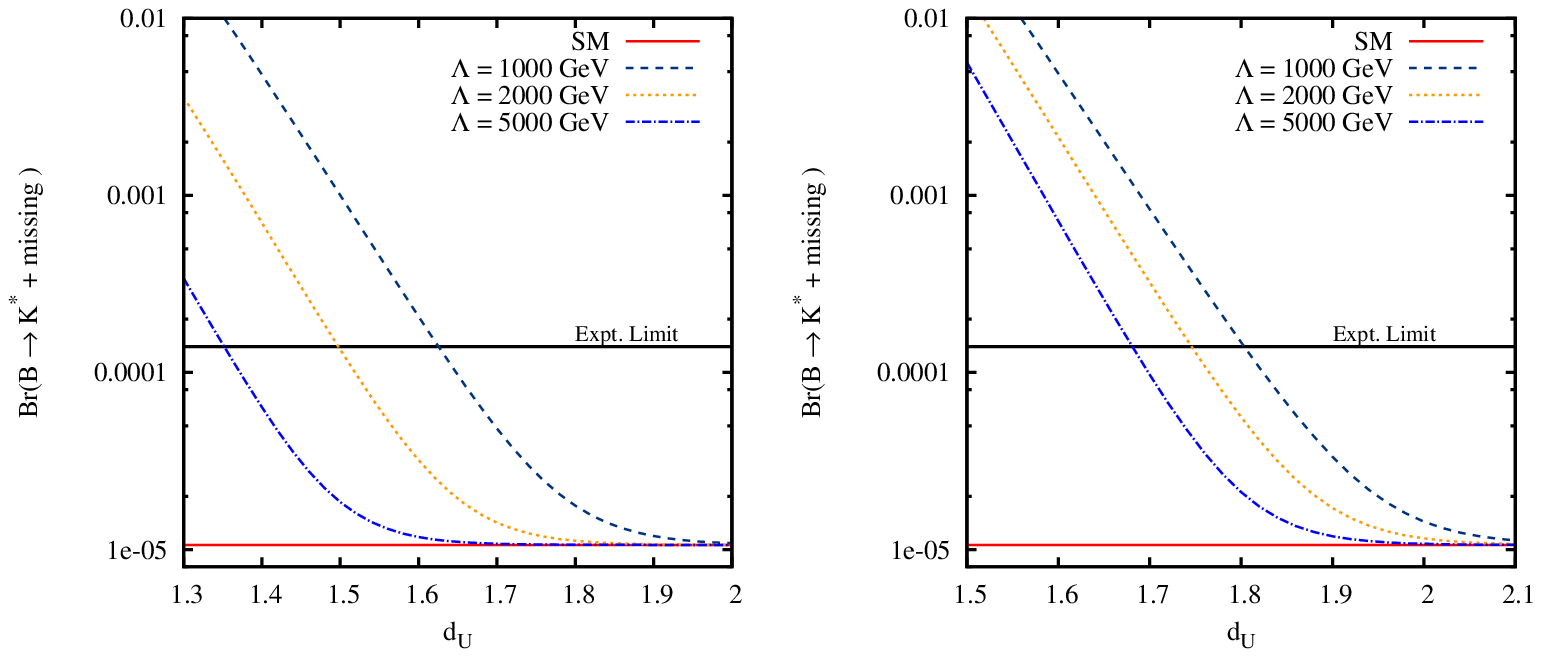,width=\textwidth} 
\caption{\it The branching ratio for $B \to K^* + \misse$ as a
  function of $\du$ for various values of $\Lambda_\unpart$. The left
  panel is for the contribution from the scalar operator, and the
  right panel is for the vector operator. The other parameters are
  $\cp = 2\times10^{-3}$ and $\cv = \ca = 10^{-5}$.}    
\label{fig:3}
\ecen
\end{figure}

 In the next set of figures, figure (\ref{fig:2}), we have tried
to analyze the constraints on the unparticle's scaling dimensions
($\du$) from different values of the cut-off scale $\lamu$. In these
plots we have used some specific values of the effective couplings
$\cs$, $\cp$, $\cv$ and $\ca$. As we can see from these figures the
branching ratio is very sensitive to the scale dimension $\du$ and
$\lamu$. In figure (\ref{fig:3}) we have shown the same plots for $B
\to K^* + \misse$. From these two figures we can observe that the
vector operators are more strongly constrained as compared to scalar
operators. The second feature is that $B \to K + \misse$ provides
better constraints than the $B \to K^* + \misse$ decay.   
 
\par
\begin{figure}[htb]
\bcen
\epsfig{file=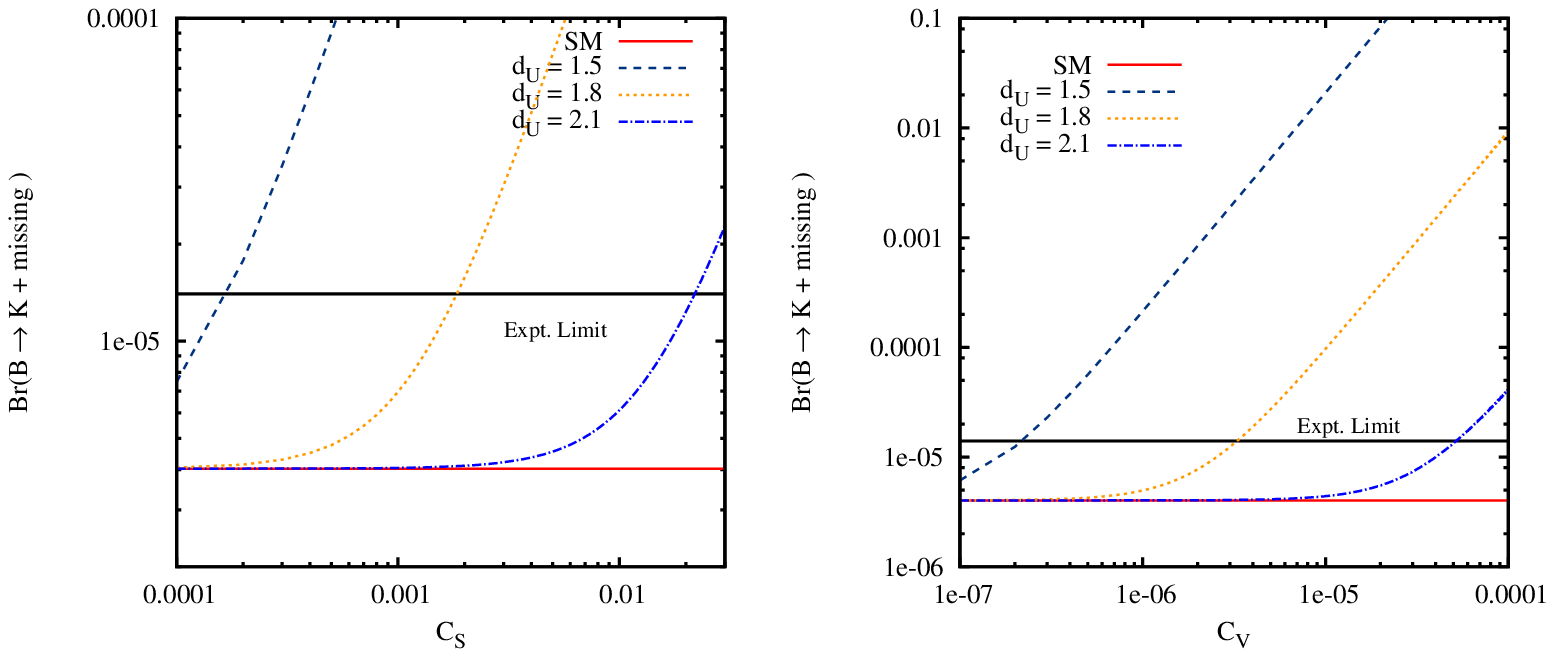,width=\textwidth} 
\caption{\it The branching ratio for $B \to K + \misse$ as a function
  of $\cs$ (left panel) and $\cv$ (right panel). The cutoff scale has
  been taken to be $\lamu = 1000$GeV.} 
\label{fig:4}
\ecen
\end{figure}

 We have next tried to estimate the limits on the allowed values
of the effective couplings, $\cs$, $\cp$, $\cv$ and $\ca$, from the
present experimental limits on the branching ratio of $B \to K(K^*) +
\misse$. Therefore, in figure (\ref{fig:4}) we have shown the
dependence of the branching ratio of $B \to K + \misse$ on $\cs$ and
$\cv$. As we can see from the expressions of the differential decay
rate for $B \to K + \misse$, given in the previous section, if we
consider the scalar (vector) operators, then the rate for this process
is only dependent on $\cs$ ($\cv$).  

\par
\begin{figure}[htb]
\bcen
\epsfig{file=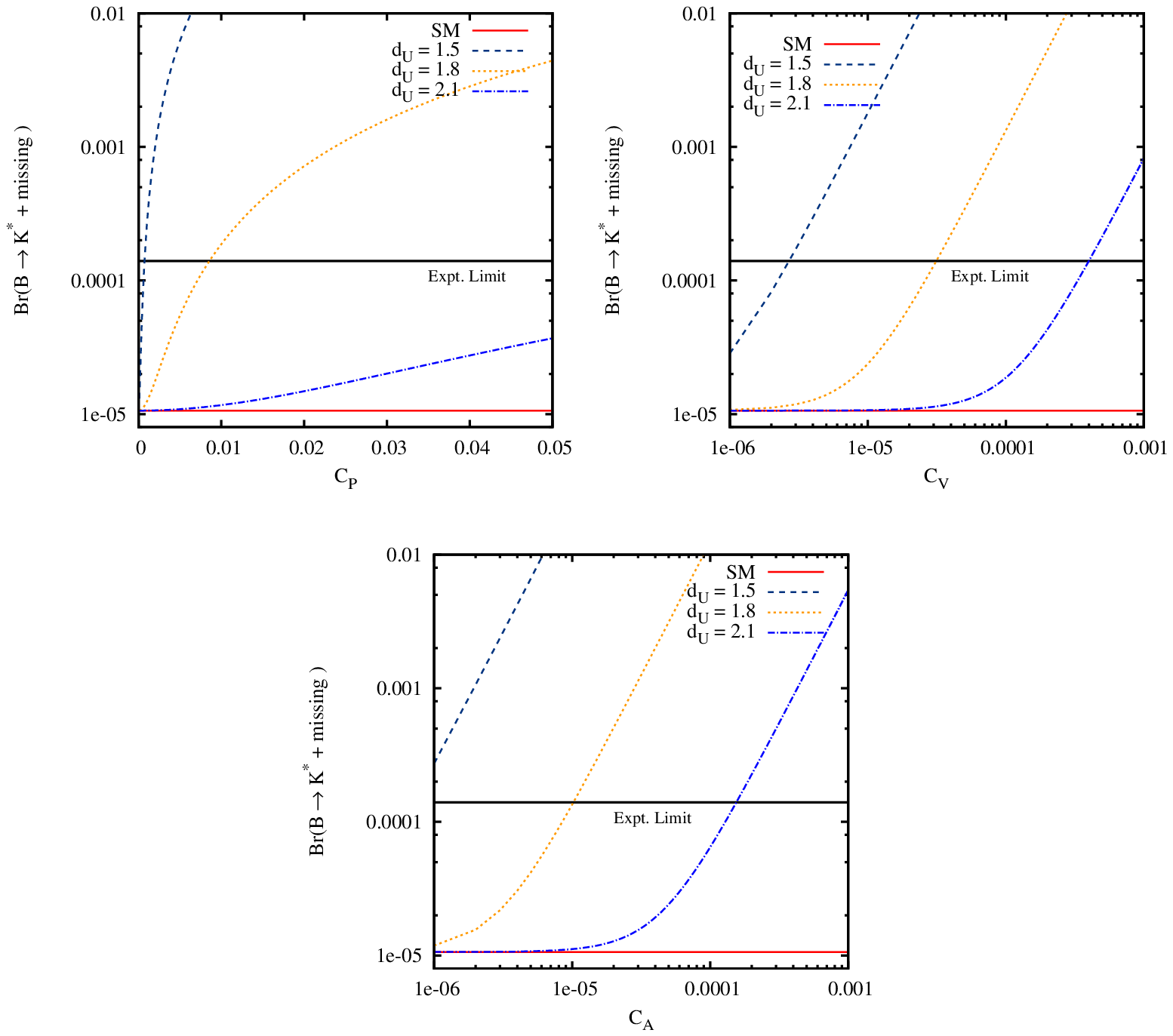,width=\textwidth} 
\caption{\it The branching ratio for $B \to K^* + \misse$ as a
  function of $\cp$ (top left panel), $\cv$ (top right panel) and
  $\ca$ (bottom panel). The cutoff scale has been taken to be $\lamu =
  1000$GeV.} 
\label{fig:5}
\ecen
\end{figure}

 Finally, in figure (\ref{fig:5}) we have shown the dependence of
the branching ratio of $B \to K^* + \misse$ on the effective
vertices. If we consider scalar operators then the rate of this
process is only dependent upon $\cp$, whereas if we consider the
vector operators then the rate can depend upon both $\cv$ and $\ca$.  

\par To re-emphasize these last few points:
\begin{itemize}
\item{} $B \to K +$ scalar unparticle operator shall constrain the parameter $\cs$,
\item{} $B \to K^* +$ scalar unparticle operator shall constrain $\cp$,
\item{} $B \to K +$ vector unparticle operator will constrain only $\cv$,
\item{} whilst $B \to K^* +$ vector unparticle operator will constrain both $\cv$ and $\ca$.
\end{itemize}

\par To conclude, in this work we have analyzed the effects of
unparticles on the missing energy signatures of rare $B$-decays. We
have tried to argue that $B \to K(K^*) + \misse$ provides very useful
constraints on the parameters of the model, where we have considered
four operators, namely the scalar, pseudo-scalar, vector and axial
vector operators. Both the modes $B \to K + \misse$ and $B \to K^* +
\misse$ are different functions of these four operators, and hence
provide independent constraints on the parameter space of the
model. The set of operators we have considered, in principal, also
contributes to meson anti-meson mixing, namely, $K - \bar{K}$, $B_d -
\bar{B}_d$, $B_s - \bar{B}_s$ and $D - \bar{D}$ mixing. Detailed
analyses of these, within the context of unparticle physics, has been
done in reference \cite{Li:2007by}. Finally, note that the constraints
provided by $B \to K(K^*) + \misse$ in some regions can be much
stronger than the ones provided by meson anti-meson mixing. 


\section*{Acknowledgement}

The work of NG was supported by JSPS under grant no. P-06043. NG would
also like to thank Yasuhiro Okada and Sukanta Dutta for the
discussions he had with them. We would also like to thank Steven
Robertson and Kai-Feng Chen for their comments regarding missing
energy signatures at $B$-factories.       

\appendix


\section{The Form Factors}

\subsection{The form factors for the $B \to K$
  transition \label{appendix:a1} }

\par The form factors for the $B \to K$ transition can be written as \cite{Ball:2004rg}: 
\beqa
\langle K(p') | \bar{s} \gamma_\mu b | B(p)\rangle & = & (p + p')_\mu f_+  + q_\mu f_- \,\,\, , \nonumber \\ 
\langle K(p') | \bar{s} \gamma_\mu \gamma_5 b | B(p)\rangle & = & 0 \,\,\, , \label{eq:ff1}
\eeqa
where $q = p - p'$. Or alternately from the light cone sum rules \cite{Ball:2004ye} as:
\beq
\langle K(p') | \bar{s} \gamma_\mu b | B(p)\rangle = \bigg\{ (p + p')_\mu - \frac{m_B^2 - m_K^2}{q^2} q_\mu \bigg\} f_+^P(q^2) + \bigg\{\frac{m_B^2 - m_K^2}{q^2} q_\mu \bigg\} f_-^P(q^2) \,\,\,  . \label{eq:ff2}
\eeq
Note that we can relate these two sets of form factors by:
\beqa
f_+ & = & f_+^P \,\,\, , \nonumber \\
f_- &=& \frac{m_B^2 - m_K^2}{q^2} \left(f_0^P - f_+^P \right) \,\,\, . \label{eq:ff3}
\eeqa
In our numerical results we have followed the parameterization of Ball and Zwicky \cite{Ball:2004ye}:
\beqa
f_0^P &=& \frac{r_2}{1 - q^2/m^2_{fit}} \,\,\, , \nonumber \\
f_+^P &=& \frac{r_1}{1 - q^2/m_1^2} + \frac{r_2}{(1 - q^2/m_1^2)^2} \,\,\, , \label{eq:ff4}
\eeqa
where the fitted parameters are given in Table \ref{table:1}.

\begin{table}[htb]
\bcen
\begin{tabular}{| c | c c c c |} \hline
          & $m_1$ & $r_1$ & $r_2$ & $m_{fit}$ \\  \hline 
$f_+^P$   & 5.41  & 0.1616 & 0.1730  &  -  \\
$f_0^P$   &  -    &   -    & 0.3302  & 5.41  \\ \hline 
\end{tabular}
\caption{{\it The parameters for the $B \to K$ form factors \cite{Ball:2004rg}.}}
\label{table:1}
\ecen
\end{table}

\subsection{The form factors for the $B \to K^*$
  transition \label{appendix:a2}} 

\par The form factors for the $B \to K^*$ transition can be written as \cite{Ball:2004rg}: 
\beqa
\langle K^*(p') | \bar{s} \gamma_\mu b | B(p)\rangle &=&
\epsilon_{\mu\nu \rho \sigma} \epsilon^\nu p^\rho p'^\sigma \frac{2
  V(q^2)}{m_B + m_{K^*}} \,\,\, , \nonumber \\ 
\langle K^*(p') | \bar{s} \gamma_\mu \gamma_5 b | B(p)\rangle &=& 
 i \epsilon_\mu (m_B + m_{K^*}) A_1(q^2) -
i (p + p')_\mu (\epsilon.q) \frac{A_2(q^2)}{m_B + m_{K^*}}  \nonumber
\\
&& \hspace{1cm} - i q_\mu (\epsilon . q) \frac{2 m_{K^*}}{q^2} 
[ A_3(q^2) - A_0(q^2)] \,\,\, , \label{eq:K*:ff1}
\eeqa
where have again defined $q = p - p'$. For this transition we have used the parameterization of reference \cite{Ball:2004rg}:
\beqa
F(q^2) &=& \frac{r_1}{1 - q^2/m_R^2}
        + \frac{r_2}{1 - q^2/m_{fit}^2} \,\,\, , ~~~~~~~~
{\rm (for \ V, A_0) }   \nonumber \\
F(q^2) &=& \frac{r_1}{1 - q^2/m_{fit}^2}
        + \frac{r_2}{(1 - q^2/m_{fit}^2)^2} \,\,\, , ~~~~~~~~
{\rm (for \ A_2) }   \nonumber \\
F(q^2) &=& \frac{r_2}{1 - q^2/m_{fit}^2} \,\,\, , ~~~~~~~~
{\rm (for \ A_1) } 
\label{eq:K*:ff2}
\eeqa
where
\beq
A_3(q^2) = \frac{m_B + m_{K^*}}{2 m_{K^*}} A_1(q^2) 
     - \frac{m_B - m_{K^*}}{2 m_{K^*}} A_2(q^2) \,\,\, .
\label{eq:K*:ff3}
\eeq
Note that the fitted parameters used in the above equations have been
given in Table \ref{table:2}. 

\begin{table}[htb]
\bcen
\begin{tabular}{| c | c c c c |} \hline
       & $r_1$ & $r_2$ & $m_{fit}^2$ & $m_R$ \\  \hline 
$V$    & 0.923 & - 0.511  & 49.40 & 5.32  \\
$A_0$  & 1.364 & - 0.99   & 36.78 & 5.63  \\ 
$A_1$  &  -    &   0.290 & 40.38  & - \\ 
$A_2$  & -0.084&   0.342  & 52.00 & - \\ \hline 
\end{tabular}
\caption{{\it The parameters for the $B \to K^*$ form factors \cite{Ball:2004rg}.}} 
\label{table:2}
\ecen
\end{table}



\begin{thebibliography}{1}


\bibitem{Buchalla:2000sk}
  G.~Buchalla, G.~Hiller and G.~Isidori,
  Phys.\ Rev.\  D {\bf 63}, 014015 (2001)
  [arXiv:hep-ph/0006136].


\bibitem{Kim:1999wa}
  C.~S.~Kim, Y.~G.~Kim and T.~Morozumi,
  Phys.\ Rev.\  D {\bf 60}, 094007 (1999)
  [arXiv:hep-ph/9905528];


\bibitem{Bird:2004ts}
  C.~Bird, P.~Jackson, R.~Kowalewski and M.~Pospelov,
  Phys.\ Rev.\ Lett.\  {\bf 93}, 201803 (2004)
  [arXiv:hep-ph/0401195] ;
  C.~Bird, R.~Kowalewski and M.~Pospelov,
  Mod.\ Phys.\ Lett.\  A {\bf 21}, 457 (2006)
  [arXiv:hep-ph/0601090].


\bibitem{Aliev:2001in}
  T.~M.~Aliev, A.~Ozpineci and M.~Savci,
  Phys.\ Lett.\  B {\bf 506}, 77 (2001)
  [arXiv:hep-ph/0101066].
%
  C.~Bobeth, A.~J.~Buras, F.~Kruger and J.~Urban,
  Nucl.\ Phys.\  B {\bf 630}, 87 (2002)
  [arXiv:hep-ph/0112305] ;
%
  J.~H.~Jeon, C.~S.~Kim, J.~Lee and C.~Yu,
  Phys.\ Lett.\  B {\bf 636}, 270 (2006)
  [arXiv:hep-ph/0602156] ;
%
  T.~M.~Aliev and C.~S.~Kim,
  Phys.\ Rev.\  D {\bf 58}, 013003 (1998)
  [arXiv:hep-ph/9710428].


\bibitem{Mahajan:2003gx}
  N.~Mahajan,
  Phys.\ Rev.\  D {\bf 68}, 034012 (2003).


\bibitem{Georgi:2007ek}
  H.~Georgi,
  arXiv:hep-ph/0703260.

\bibitem{Banks:1981nn}
  T.~Banks and A.~Zaks,
  Nucl.\ Phys.\  B {\bf 196}, 189 (1982).

\bibitem{Georgi:2007si}
  H.~Georgi,
  arXiv:0704.2457 [hep-ph].

\bibitem{Luo:2007bq}
  M.~Luo and G.~Zhu,
  arXiv:0704.3532 [hep-ph] ;
%
  C.~H.~Chen and C.~Q.~Geng,
  arXiv:0705.0689 [hep-ph].


\bibitem{Cheung:2007ue}
  K.~Cheung, W.~Y.~Keung and T.~C.~Yuan,
  arXiv:0704.2588 [hep-ph].


\bibitem{Liao:2007bx}
  Y.~Liao,
  arXiv:0705.0837 [hep-ph].

\bibitem{Li:2007by}
  X.~Q.~Li and Z.~T.~Wei,
  arXiv:0705.1821 [hep-ph].


\bibitem{Choudhury:2007js}
  D.~Choudhury, D.~K.~Ghosh and Mamta,
  arXiv:0705.3637 [hep-ph] ; 
%
  T.~M.~Aliev, A.~S.~Cornell and N.~Gaur,
  arXiv:0705.1326 [hep-ph] ;
%
  C.~D.~Lu, W.~Wang and Y.~M.~Wang,
  arXiv:0705.2909 [hep-ph].


\bibitem{Stephanov:2007ry}
  M.~A.~Stephanov,
  arXiv:0705.3049 [hep-ph].


\bibitem{Fox:2007sy}
  P.~J.~Fox, A.~Rajaraman and Y.~Shirman,
  arXiv:0705.3092 [hep-ph].
%
  N.~Greiner,
  arXiv:0705.3518 [hep-ph] ;
%
  S.~L.~Chen and X.~G.~He,
  arXiv:0705.3946 [hep-ph].

\bibitem{Davoudiasl:2007jr}
  H.~Davoudiasl,
  arXiv:0705.3636 [hep-ph].


\bibitem{explimits}
Kai-Feng Chen [Belle Collaboration] Talk given at FPCP07, Slovenia,
May 12-16, 2007. 

\bibitem{Aubert:2004ws}
  B.~Aubert {\it et al.}  [BABAR Collaboration],
  Phys.\ Rev.\ Lett.\  {\bf 94} (2005) 101801
  [arXiv:hep-ex/0411061].


\bibitem{Ball:2004rg}
  P.~Ball and R.~Zwicky,
  Phys.\ Rev.\  D {\bf 71}, 014029 (2005)
  [arXiv:hep-ph/0412079].

\bibitem{Ball:2004ye}
  P.~Ball and R.~Zwicky,
  Phys.\ Rev.\  D {\bf 71}, 014015 (2005)
  [arXiv:hep-ph/0406232].

\end{thebibliography}
\end{document}